\documentstyle[preprint,tighten,prl,aps,floats,psfig]{revtex}

\begin{document}

\preprint{IFUP-TH/2000-02}

\newcommand{\N}{\hbox{{\rm I}\kern-.2em\hbox{\rm N}}}

\title{
The critical equation of state of three-dimensional $XY$  systems
}

\author{Massimo Campostrini$\,^1$, Andrea Pelissetto$\,^2$, 
        Paolo Rossi$\,^1$, Ettore Vicari$\,^1$ }
\address{$^1$ Dipartimento di Fisica dell'Universit\`a di Pisa 
and I.N.F.N., I-56126 Pisa, Italy}
\address{$^2$ Dipartimento di Fisica dell'Universit\`a di Roma I
and I.N.F.N., I-00185 Roma, Italy \\
{\bf e-mail: \rm
{\tt campo@mailbox.difi.unipi.it}, 
{\tt rossi@mailbox.difi.unipi.it},
{\tt Andrea.Pelissetto@roma1.infn.it},
{\tt vicari@mailbox.difi.unipi.it}
}}
\date{\today}

\maketitle

\begin{abstract}
We address the problem of determining the critical equation of state of 
three-dimensional $XY$ systems. 
For this purpose we first consider
the small-field expansion of the effective potential
(Helmholtz free energy) in the high-temperature phase. We compute
the first few nontrivial zero-momentum $n$-point renormalized couplings,
which parametrize such expansion, by analyzing the high-temperature
expansion of an improved lattice Hamiltonian with suppressed
leading scaling corrections. 

These results are then used to construct parametric representations of
the critical equation of state which are valid in the whole critical
regime, satisfy the correct analytic properties (Griffith's
analyticity), and take into account the Goldstone singularities at the
coexistence curve.  A systematic approximation scheme is introduced,
which is limited essentially by the number of known terms in the
small-field expansion of the effective potential.

From our approximate representations of the  equation of state,
we derive estimates of universal ratios of amplitudes.
For the specific-heat amplitude ratio  
we obtain $A^+/A^-=1.055(3)$, to be compared with the
best experimental estimate  $A^+/A^-=1.054(1)$.

\end{abstract}

\vskip1.5pc
\bgroup\small
\leftskip=0.10753\textwidth \rightskip\leftskip
\par\egroup
\vskip1.5pc


\section{Introduction}

In the theory of critical phenomena, continuous phase transitions can
be classified into universality classes determined only by a few basic
properties characterizing the system, such as the space
dimensionality, the range of interaction, the number of components and
the symmetry of the order parameter.  The renormalization-group theory
predicts that, within a given universality class, the critical
exponents and the scaling functions are the same for all systems.
Here we consider the $XY$ universality class, which is characterized
by a two-component order parameter and effective short-range
interactions.  The lattice spin model described by the Hamiltonian
\begin{equation}
{\cal H}_L = - J \sum_{<ij>} \vec{s}_i\cdot\vec{s}_j 
+ \sum_i \vec{h}_i\cdot \vec{s}_i ,
\label{lattspin}
\end{equation}
where $\vec{s}_i$ is a two-component spin satisfying
$\vec{s}_i\cdot\vec{s}_i=1$, is one of the systems belonging to the
$XY$ universality class.  It may be viewed as a magnetic system with
easy-plane anisotropy, in which the magnetization plays the role of
order parameter and the spins are coupled to an external magnetic
field $h$.

The superfluid transition of $^4$He, occurring along the
$\lambda$-line $T_\lambda(P)$ (where $P$ is the pressure), belongs to
the three-dimensional $XY$ universality class. Its order parameter is
related to the complex quantum amplitude of helium atoms.  It provides
an exceptional opportunity for an experimental test of the
renormalization-group predictions, thanks to the weakness of the
singularity in the compressibility of the fluid and to the purity of
the sample. Moreover, experiments in a microgravity environment lead
to a reduction of the gravity-induced broadening of the transition.
Recently a Space Shuttle experiment \cite{LSNCI-96} performed a very
precise measurement of the heat capacity of liquid helium to within 2
nK from the $\lambda$-transition obtaining an extremely accurate
estimate of the exponent $\alpha$ and of the ratio $A^+/A^-$ of the
specific-heat amplitudes:
\begin{equation}
\alpha = - 0.01285(38),\qquad\qquad
A^+/A^- = 1.054(1). \label{expspace}
\end{equation}
These results represent a challenge for theorists because the accuracy
of the test of the renormalization-group prediction is now limited by
the precision of the theoretical calculations.  We mention the best
available theoretical estimates for $\alpha$: $\alpha=-0.0150(17)$
obtained using high-temperature expansion techniques \cite{CPRV-00},
$\alpha=-0.0169(33)$ from Monte Carlo simulations using finite-size
scaling techniques \cite{HT-99}, $\alpha=-0.011(4)$ from field theory
~\cite{GZ-98}.  The close agreement with the experimental data clearly
supports the standard renormalization-group description of the
$\lambda$-transition \cite{footnote_singularity}.

In this paper we address the problem of determining the critical
equation of state characterizing the $XY$ universality class.  The
critical equation of state relates the thermodynamical quantities in
the neighborhood of the critical temperature, in both phases.  It is
usually written in the form (see e.g. Ref.~\cite{ZJbook})
\begin{equation}
\vec{H} = \vec{M} M^{\delta-1} f(x), \qquad\qquad x \propto t M^{-1/\beta},
\label{eqstfx}
\end{equation}
where $f(x)$ is a universal scaling function
(normalized in such a way that $f(-1)=0$ and $f(0)=1$).
The universal ratios of amplitudes involving quantities
defined at zero-momentum (i.e.\ integrated in the volume), such as
specific heat, magnetic susceptibility, etc..., can be obtained from 
the scaling function $f(x)$. 

It should be noted that, for the $\lambda$-transition in ${}^4$He, 
Eq.\ (\ref{eqstfx})  is not directly related to the conventional equation
of state that relates temperature and pressure. Moreover, in
this case, the field $\vec H$ does not correspond to an 
experimentally accessible external field, so that the function 
appearing in Eq.\ (\ref{eqstfx}) cannot be determined directly in experiments.
The physically interesting quantities are universal amplitude ratios of 
quantities formally defined at zero external field.

As our starting point for the determination of the critical equation
of state, we compute the first few nontrivial coefficients of the
small-field expansion of the effective potential (Helmholtz free
energy) in the high-temperature phase.  For this purpose, we analyze
the high-temperature expansion of an improved lattice Hamiltonian with
suppressed leading scaling corrections~\cite{CFN-82,CPRV-99}.  If the
leading non-analytic scaling corrections are no longer present, one
expects a faster convergence, and therefore an improved
high-temperature expansion (IHT) whose analysis leads to more precise
and reliable estimates.  We consider a simple cubic lattice and the
$\phi^4$ Hamiltonian
\begin{equation}
{\cal H}=\, - \beta\sum_{\langle x,y\rangle} {\vec\phi}_x\cdot{\vec\phi}_y +\, 
         \sum_x \left[ {\vec\phi}_x^2 + \lambda ({\vec\phi}_x^2 - 1)^2\right],
\label{Hamiltonian}
\end{equation}
where $\langle x,y\rangle$ labels a lattice link, and ${\vec\phi}_x$
is a real two-component vector defined on lattice sites.  The value of
$\lambda$ at which the leading corrections vanish has been determined
by Monte Carlo simulations using finite-size techniques~\cite{HT-99},
obtaining $\lambda^* = 2.10(6)$.  In Ref.~\cite{CPRV-00} we have
already considered the high-temperature expansion (to 20th order) of
the improved $\phi^4$ Hamiltonian (\ref{Hamiltonian}) for the
determination of the critical exponents, achieving a substantial
improvement with respect to previous theoretical estimates.  IHT
expansions have also been considered for Ising-like
systems~\cite{CPRV-99}, obtaining accurate determinations of the
critical exponents, of the small-field expansion of the effective
potential, and of the small-momentum behavior of the two-point
function.

We use the small-field expansion of the effective potential in the
high-temperature phase to determine approximate representations of the
equation of state that are valid in the whole critical regime.  To
reach the coexistence curve ($t<0$) from the high-temperature phase
($t>0$), an analytic continuation in the complex
$t$-plane~\cite{ZJbook,GZ-97} is required.  For this purpose we use
parametric representations \cite{Schofield-69,SLH-69,Josephson-69},
which implement in a rather simple way the known analytic properties
of the equation of state (Griffith's analyticity).  This approach was
successfully applied to the Ising model, for which one can construct a
systematic approximation scheme based on polynomial parametric
representations~\cite{GZ-97} and on a global stationarity
condition~\cite{CPRV-99}.  This leads to an accurate determination of
the critical equation of state and of the universal ratios of
amplitudes that can be extracted from it~\cite{GZ-97,GZ-98,CPRV-99}.
$XY$ systems, in which the phase transition is related to the breaking
of the continuous symmetry O(2), present a new important feature with
respect to Ising-like systems: the Goldstone singularities at the
coexistence curve.  General arguments predict that at the coexistence
curve ($t<0$ and $H\rightarrow 0$) the transverse and longitudinal
magnetic susceptibilities behave respectively as
\begin{equation}
\chi_T = {M\over H}  ,\qquad\qquad
\chi_L = {\partial M\over \partial H} \sim H^{d/2-2}.
\label{chitl}
\end{equation}
In our analysis we will consider polynomial parametric representations 
that have the correct singular behavior at the coexistence curve.

The most important result of the present paper, from the point of view
of comparison with experiments, is the specific-heat amplitude ratio;
our final estimate
\begin{equation}
A^+/A^-=1.055(3)
\end{equation}
is perfectly consistent with the experimental estimate (\ref{expspace}),
although not as precise.

The paper is organized as follows.  In Sec.~\ref{sec2} we study the
small-field expansion of the effective potential (Helmholtz free
energy). We present estimates of the first few nontrivial coefficients
of such expansion obtained by analyzing the IHT series.  The results
are then compared with other theoretical estimates.  In
Sec.~\ref{sec3}, using as input parameters the critical exponents and
the known coefficients of the small-field expansion of the effective
potential, we construct approximate representations of the critical
equation of state.  We obtain new estimates for many universal
amplitude ratios.  These results are then compared with experimental
and other theoretical estimates.

\section{The effective potential in the high-temperature phase}
\label{sec2}

\subsection{Small-field expansion of the effective potential in the
high-temperature phase}
\label{sec2a}

The effective potential (Helmholtz free energy) is related to the
(Gibbs) free energy of the model.  If $\vec{M}\equiv\langle
\vec{\phi}\rangle$ is the magnetization and $\vec{H}$ the magnetic field, one
defines
\begin{equation}
{\cal F} (M) = \vec{M} \cdot \vec{H} - {1\over V} \log Z(H),
\end{equation}
where $Z(H)$ is the partition function and the dependence on the
temperature is always understood in the notation.

In the high-temperature phase the effective potential admits an
expansion around $M=0$:
\begin{equation}
\Delta {\cal F} \equiv {\cal F} (M) - {\cal F} (0) = 
\sum_{j=1}^\infty {1\over (2j)!} a_{2j} M^{2j}.
\end{equation}
This expansion can be rewritten in terms of a renormalized
magnetization $\varphi$
\begin{equation}
\Delta {\cal F}= {1\over 2} m^2\varphi^2 + 
\sum_{j=2} m^{3-j} {1\over (2j)!} g_{2j} \varphi^{2j} 
\label{freeeng}
\end{equation}
where
\begin{equation}
\varphi^2 = {\xi(t,H=0)^2 M(t,H)^2\over \chi(t,H=0)} ,
\end{equation}
$t$ is the reduced temperature,
$\chi$ and $\xi$ are respectively the magnetic susceptibility
and the second-moment correlation length
\begin{eqnarray}
&&\chi = \sum_x \langle \phi_\alpha(0)\phi_\alpha(x) \rangle ,\label{chixi}\\
&&\xi = {1\over 6\chi} 
\sum_x x^2 \langle \phi_\alpha(0)\phi_\alpha(x) \rangle ,\nonumber
\end{eqnarray}
and $m\equiv 1/\xi$.  In field theory $\varphi$ is the expectation
value of the zero-momentum renormalized field.  The zero-momentum
$2j$-point renormalized constants $g_{2j}$ approach universal
constants (which we indicate with the same symbol) for $t\to 0$.  By
performing a further rescaling
\begin{equation}
\varphi = {m^{1/2}\over\sqrt{g_4}} z
\label{defzeta}
\end{equation} 
in Eq.\ (\ref{freeeng}), 
the free energy can be  written as
\begin{equation}
\Delta {\cal F} = {m^3\over g_4}A(z),
\label{dAZ}
\end{equation}
where
\begin{equation}
A(z) =   
{1\over 2} z^2 + {1\over 4!} z^4 + \sum_{j=3} {1\over (2j)!} r_{2j} z^{2j},
\label{AZ}
\end{equation}
and
\begin{equation}
r_{2j} = {g_{2j}\over g_4^{j-1}} \qquad\qquad j\geq 3.
\label{r2j}
\end{equation}
One can show that $z\propto t^{-\beta} M$, and
that the equation of state can be written in the form
\begin{equation}
H\propto t^{\beta\delta} {\partial A(z)\over \partial z}.
\label{eqa}
\end{equation}

\subsection{Zero-momentum renormalized couplings by IHT expansion}
\label{sec2b}

To compute the high-temperature  series of the four-point coupling
$g_4$ and  of the effective-potential parameters
$r_{2j}$, we rewrite them in terms of the
zero-momentum connected $2j$-point Green's functions $\chi_{2j}$
\begin{equation}
\chi_{2j} = 
\sum_{x_2,...,x_{2j}}\langle \phi_{\alpha_1}(0) \phi_{\alpha_1}(x_2)
...\phi_{\alpha_j}(x_{2j-1}) \phi_{\alpha_j}(x_{2j})\rangle_c
\end{equation}
($\chi = \chi_2$). For generic $N$-vector models we have
\begin{equation}
g_4 = - {3N\over N+2} {\chi_4\over \chi_2^2 \xi^3}
\label{grdef}
\end{equation}
and
\begin{eqnarray}
r_6 =&& 10 - {5(N+2)\over 3(N+4)}{\chi_6\chi_2\over \chi_4^2}, 
        \label{r2jgreen}\\
r_8 =&& 280 - {280 (N+2)\over 3(N+4)}{\chi_6\chi_2\over \chi_4^2} 
+{35(N+2)^2\over 9(N+4)(N+6)}{\chi_8\chi_2^2\over \chi_4^3},\nonumber\\
r_{10} =&& 
15400  
-{7700  (N + 2)\over (N + 4)} {\chi_6 \chi_2\over \chi_4^2}        
+{ 350  (N + 2)^2\over(N + 4)^2} {\chi_6^2 \chi_2^2\over \chi_4^4} 
\nonumber \\ 
 && +{1400 (N + 2)^2\over 3(N + 4) (N + 6)} {\chi_8 \chi_2^2\over \chi_4^3} 
-{35 (N + 2)^3\over 3(N + 4) (N + 6) (N + 8)} 
                   {\chi_{10} \chi_2^3\over \chi_4^4}.
\nonumber
\end{eqnarray}
The formulae relevant for the $XY$ universality class are obtained
setting $N=2$.

Using the $\phi^4$ lattice Hamiltonian (\ref{Hamiltonian}), we have
calculated $\chi$ and 
$m_2\equiv \sum_x x^2 \langle \phi(0) \phi(x) \rangle$ to 20th order,
$\chi_4$ to 18th order, $\chi_6$ to 17th order, $\chi_8$ to 16th
order, and $\chi_{10}$ to 15th order, for generic values of $\lambda$.
The IHT expansion, i.e.\ with suppressed leading scaling corrections,
is achieved for $\lambda=2.10(6)$~\cite{HT-99}.  In
Table~\ref{HTexpansions} we report the series of $m_2$, $\chi$,
$\chi_4$, $\chi_6$, $\chi_8$ and $\chi_{10}$ for $\lambda=2.10$.
Using Eqs.\ (\ref{grdef}) and (\ref{r2jgreen}) one can obtain the HT
series necessary for the determination of $g_4$ and $r_{2j}$.  We
analyzed the series using the same procedure applied to the improved
high-temperature expansions of Ising-like systems in
Ref.~\cite{CPRV-99}. In order to estimate the fixed-point value of
$g_4$ and $r_{2j}$, we considered Pad\'e, Dlog-Pad\'e and first-order
integral approximants of the series in $\beta$ for $\lambda=2.10$, and
evaluated them at $\beta_c$.  We refer to Ref.~\cite{CPRV-99} for the
details of the analysis.  Our estimates are
\begin{eqnarray}
&&g_4 = 21.05(3+3),\\
&&\bar{g}\equiv  {5 \over 24\pi}g_4 = 1.396(2+2),\\
&& r_6 = 1.951(11+3) ,\\
&& r_8 = 1.36(6+3) .
\end{eqnarray}
We quote two errors: the first one is related to the spread of the
approximants, while the second one gives the variation of the estimate
when $\lambda$ varies between 2.04 and 2.16.  In addition, we obtained
a rough estimate of $r_{10}$, i.e.\ $r_{10}=-13(7)$.  For comparison,
we anticipate that the analysis of the critical equation of state
using approximate parametric representations will lead to the estimate
$r_{10}=-10(3)$.  From the estimates of $g_4^*$ and $r_{2j}$ one can
obtain corresponding estimates for the zero-momentum renormalized
couplings, $g_{2j}= r_{2j} g_4^{j-1}$ with $j>2$.

\begin{table}[tbp]
\caption{Coefficients of the high-temperature expansion of $m_2$,
$\chi$, $\chi_4$, $\chi_6$, $\chi_8$, $\chi_{10}$. They have been
obtained using the Hamiltonian (\protect\ref{Hamiltonian}) with
$\lambda=2.10$.
}
\label{HTexpansions}
\begin{tabular}{cccc}
$i$& \multicolumn{1}{c}{$m_2$} &
     \multicolumn{1}{c}{$\chi$} &
     \multicolumn{1}{c}{$\chi_4$} \\ 
\hline
0  & 0  &                          0.82305062235187783163838  &  
  $-$0.18361621025768068492172 \\ 
1  & 1.01611849043072013077966  &  2.03223698086144026155932  &  
  $-$1.81350523351772810305606 \\ 
2  & 5.01790173559352893587925  &  4.41339999108761956819390  &  
  $-$10.2403328491873632045378 \\ 
3  & 17.1372887645210686867092  &  9.49465465075720590174531  &  
  $-$44.8928073800743626690682 \\ 
4  & 50.5729609164717064337798  &  19.8432084757850795073880  &  
  $-$168.753294310535712758532 \\ 
5  & 136.975729153185634797939  &  41.3105316077169269675024  &  
  $-$572.899325351072412028155 \\ 
6  & 351.547955204015543873533  &  84.8657665346587543517940  &  
  $-$1806.05498365521624417337 \\ 
7  & 867.882609960590948513531  &  173.919877080583932668034  &  
  $-$5384.52486767607368575700 \\ 
8  & 2082.72988964511892008013  &  353.805866554914925628451  &  
  $-$15362.4317708302607812112 \\ 
9  & 4887.89348922134368695104  &  718.544573182206654200415  &  
  $-$42306.4841196583464482445 \\ 
10 & 11271.6038365544561486584  &  1452.55696588171500306095  &  
  $-$113150.461045583807761395 \\ 
11 & 25618.1126339930289598217  &  2932.78243299692312158659  &  
  $-$295305.926162214852952064 \\ 
12 & 57532.0447298988841107389  &  5902.70244465199387875428  &  
  $-$754781.109517718395587745 \\ 
13 & 127889.749773305781861722  &  11869.0080781568501056056  &  
  $-$1894806.32404641680977111 \\ 
14 & 281828.971329949201651321  &  23811.0622160012068230227  &  
  $-$4682769.64431772632678941 \\ 
15 & 616366.979493029224503143  &  47733.2907485305100325248  &  
  $-$11414677.5171522019349233 \\ 
16 & 1339133.76964227829126980  &  95522.7109197203117005540  &  
  $-$27486804.4332023911210055 \\ 
17 & 2892420.77412897864771425  &  191043.098791148001757550  &  
  $-$65472437.4333935066305991 \\ 
18 & 6215040.60947250914305650  &  381560.204019249641798944  &  
  $-$154435238.761288809747194 \\ 
19 & 13292241.3642748890173429  &  761691.955581673385411208  &  
      \\ 
20 & 28309542.5693146536002667  &  1518865.25572011244933658  &  
      \\ 
\hline\hline
$i$& \multicolumn{1}{c}{$\chi_6$} &
     \multicolumn{1}{c}{$\chi_8$} &
     \multicolumn{1}{c}{$\chi_{10}$} \\
\hline
0  & 0.41651792496748423277383  &  
  $-$2.07276872395008612032057  &  17.8224779282297038884825 \\ 
1  & 8.19357083273717982769144  &  
  $-$66.6409389277329689706620  &  845.247991215093077905035 \\ 
2  & 81.6954441314190452820070  &  
  $-$1011.50102952099242860043  &  17983.4651630934745766842 \\ 
3  & 577.734561420010600288632  &  
  $-$10327.4974814820902607880  &  248053.171125799025029980 \\ 
4  & 3288.50395189429008814810  &  
  $-$81557.4477403308614541401  &  2572480.03564061669203194 \\ 
5  & 16096.6021109696857145564  &  
  $-$536525.345841149579660234  &  21718902.5365937343351130 \\ 
6  & 70442.2092664445625747715  &  
  $-$3075097.70592321924332973  &  156737903.631432308532083 \\ 
7  & 282678.282078376456985396  &  
  $-$15817337.6236306593687791  &  998806344.593851737549288 \\ 
8  & 1058442.95252211500336034  &  
  $-$74543191.4965166803297285  &  5750852421.81974869997198 \\ 
9  & 3744677.84665924694446919  &  
  $-$326778866.937610399640322  &  30428100948.8992362353297 \\ 
10 & 12635750.8707582942375013  &  
  $-$1347823157.99877343602884  &  149865516252.140441504474 \\ 
11 & 40959287.0765118834279901  &  
  $-$5277067971.27466652475816  &  694034037396.324618345020 \\ 
12 & 128268696.512235415964465  &  
  $-$19750695382.3283907899788  &  3046464352917.44305602911 \\ 
13 & 389824946.322878356697650  &  
  $-$71066114622.3163753581174  &  12757807364908.6407194560 \\ 
14 & 1153973293.83499223649654  &  
  $-$246972985495.952557381976  &  51244942192011.6695801488 \\ 
15 & 3337461787.55665279189948  &  
  $-$832179913942.254869433972  &  198319718730601.629272462 \\ 
16 & 9454326701.58705373223814  &  
  $-$2727546459564.59095483523  &   \\ 
17 & 26288539415.6469228911099  &  
       &   \\ 
\end{tabular}
\end{table}

Table~\ref{summarygj} compares our results (denoted by IHT) with the
estimates obtained using other approaches, such as the
high-temperature expansion (HT) of the standard lattice spin model
(\ref{lattspin})~\cite{BC-98,PV-98-gr,Reisz-95}, field-theoretic
methods based on the fixed-dimension $d=3$
$g$-expansion~\cite{GZ-98,SOUK-99,MN-91} and on the
$\epsilon$-expansion~\cite{PV-98-gr,PV-98-ef,PV-00}.  The
fixed-dimension field-theoretic estimates of $g_4$ have been obtained
from the zero of the Callan-Symanzik $\beta$-function, whose expansion
is known to six loops~\cite{BNGM-77}.  In the same framework $g_6$ and
$g_8$ have been estimated from the analysis of the corresponding four-
and three- loop series respectively~\cite{SOUK-99}.  The authors of
Ref.~\cite{SOUK-99} argue that the uncertainty on their estimate of
$g_6$ is approximately 0.3\%, while they consider their value for
$g_8$ much less accurate.  The $\epsilon$-expansion estimates have
been obtained from constrained analyses of the four-loop series of
$g_4$ and the three-loop series of $r_{2j}$.

\begin{table}
\caption{
Estimates of $\bar{g}\equiv 5 g_4/(24\pi)$,
$r_{6}$, and $r_8$.}
\label{summarygj}
\begin{tabular}{ccccc}
\multicolumn{1}{c}{}&
\multicolumn{1}{c}{IHT}&
\multicolumn{1}{c}{HT}&
\multicolumn{1}{c}{$d=3\;\;g$-exp.}&
\multicolumn{1}{c}{$\epsilon$-exp.}\\
\tableline \hline
$\bar{g}$ & 1.396(4)  & 1.411(8),$\;$ 1.406(8) \cite{BC-98} & 1.403(3) \cite{GZ-98} & 1.425(24) \cite{PV-00,PV-98-gr} \\
            &           & 1.415(11)\cite{PV-98-gr} & 1.40 \cite{MN-91} & \\\hline

$r_6$       & 1.951(14) & 2.2(6) \cite{Reisz-95}  &  1.967 \cite{SOUK-99} & 1.969(12) \cite{PV-00,PV-98-ef} \\
 
$r_8$       & 1.36(9)   &                         &  1.641 \cite{SOUK-99} & 2.1(9) \cite{PV-00,PV-98-ef} \\
\end{tabular}
\end{table}

\section{The critical equation of state}
\label{sec3}

\subsection{Analytic properties of the scaling equation of state}
\label{sec3a}

From the analysis of the IHT series we have obtained the first few
non-trivial terms of the small-field expansion of the effective
potential in the high-temperature phase. This provides corresponding
information for the equation of state
\begin{equation}
H\propto t^{\beta\delta} F(z),
\label{eqa2}
\end{equation}
where $z \propto M t ^{-\beta}$ and, using Eq.\ (\ref{eqa}), 
\begin{equation}
F(z)={\partial A(z)\over \partial z} = 
z + \case{1}{6}z^3 + \sum_{m=3} F_{2m-1} z^{2m-1}
\label{Fzdef}
\end{equation}
with 
\begin{equation}
F_{2m-1} = {1\over (2m-1)!} r_{2m}.
\end{equation}
The function $H(M,t)$ representing the external field in the critical
equation of state (\ref{eqa2}) satisfies Griffith's analyticity: it is
regular at $M=0$ for $t>0$ fixed and at $t=0$ for $M>0$ fixed.  The
first region corresponds to small $z$ in Eq.\ (\ref{eqa2}), while the
second is related to large $z$, where $F(z)$ has an expansion of the form
\begin{equation}
F(z) = z^\delta \sum_{n=0} F^{\infty}_n z^{-n/\beta}.
\label{fasymp}
\end{equation}
To reach the coexistence curve, i.e.\ $t<0$ and $H=0$, one should
perform an analytic continuation in the complex $t$-plane
\cite{ZJbook,GZ-97}.  The spontaneous magnetization is related to
the complex zero $z_0$ of $F(z)$.  Therefore, the description of the
coexistence curve is related to the behavior of $F(z)$ in the
neighbourhood of $z_0$.  

\subsection{Goldstone singularities at the coexistence curve}
\label{sec3b}

The physics of the broken phase of $N$-vector models
(including $XY$ systems which correspond to $N=2$)  
is very different from that of the Ising model, because
of the presence of Goldstone modes at the coexistence curve.
The singularity of $\chi_L$ for $t<0$ and $H\to0$
is governed by the zero-temperature infrared-stable fixed
point~\cite{BW-73,BZ-76,Lawrie-81}.  This leads to the prediction 
\begin{equation}
f(x) \approx  c_f \,(1+x)^{2/(d-2)} \qquad\qquad {\rm for}
\qquad x\rightarrow -1,
\label{fxcc} 
\end{equation}
where $x\propto t M^{-1/\beta}$ and $f(x)$ is the scaling function
introduced in Eq.~(\ref{eqstfx}) (as usual, $x=-1$ corresponds to the
coexistence curve).  This behavior at the coexistence curve has been
verified in the framework of the large-$N$ expansion to $O(1/N)$
(i.e.\ next-to-leading order) \cite{BW-73,cflargeN}.

The nature of the corrections to the behaviour (\ref{fxcc}) is less
clear. Setting $\omega=1+x$ and $y=H M^{-\delta}$, it has been
conjectured that $\omega$ has the form of a double expansion in powers
of $y$ and $y^{(d-2)/2}$ near the coexistence
curve~\cite{WZ-75,SH-78,Lawrie-81}, i.e.\ for $y\to0$
\begin{equation}
\omega\equiv 1+x = c_1 y + c_2 y^{1-\epsilon/2} + d_1 y^2 + 
       d_2 y^{2-\epsilon/2} + d_3 y^{2-\epsilon} + \ldots
\label{expcoex1}
\end{equation}
where $\epsilon = 4 - d$.  This expansion has been derived essentially
from an $\epsilon$-expansion analysis~\cite{epsexp}.  Note that in
three dimensions this conjecture predicts an expansion in powers of
$y^{1/2}$, or equivalently an expansion of $f(x)$ in powers of
$\omega$ for $\omega\to 0$.

The asymptotic expansion  of the $d$-dimensional  equation of state at
the coexistence curve has been computed  analytically in the framework
of the large-$N$ expansion~\cite{PV-00},  using the $O(1/N)$  formulae
reported in  Ref.~\cite{BW-73}.    It turns  out  that  the  expansion
(\ref{expcoex1})  does not strictly hold  for  values of the dimension
$d$ such that
\begin{equation}
2 < d = 2 + {2 m\over n} < 4, \qquad {\rm for}\quad 
0< m < n,\quad  m,n\in\ \N.
\label{speciald}
\end{equation}
In particular, in three dimensions one finds~\cite{PV-99}
\begin{equation}
f(x) = \omega^2\left[1 + {1\over N}\left(
  f_1(\omega) + \log \omega f_2(\omega) \right)
    + O(N^{-2})\right] ,
\label{fx3d}
\end{equation}
where the functions $f_1(\omega)$ and $f_2(\omega)$ have a regular
expansion in powers of $\omega$. Moreover,
\begin{equation}
f_2(\omega)=O(\omega^2),
\end{equation}
so that the logarithms affect the power expansion only at the
next-next-to-leading order.  A possible interpretation of the
large-$N$ analysis is that the expansion (\ref{fx3d}) holds for all
values of $N$, so that Eq.\ (\ref{expcoex1}) is not correct due to the
presence of logarithms.  The reason of their appearance is however
unclear.  Neverthless, it does not contradict the conjecture that the
behavior near the coexistence curve is controlled by the
zero-temperature infrared-stable Gaussian fixed point.  In this case
logarithms would not be unexpected, as they usually appear in the
reduced-temperature asymptotic expansion around Gaussian fixed points
(see e.g. Ref.~\cite{BB-85}).

\subsection{Parametric representations}
\label{sec3c}

In order to obtain a representation of the critical equation of state
that is valid in the whole critical region, one may use parametric
representations, which implement in a simple way all scaling and
analytic properties\cite{Schofield-69,SLH-69,Josephson-69}.  One may
parametrize $M$ and $t$ in terms of $R$ and $\theta$ according to
\begin{eqnarray}
M &=& m_0 R^\beta m(\theta) ,\label{parrep} \\
t &=& R(1-\theta^2), \nonumber \\
H &=& h_0 R^{\beta\delta}h(\theta),\nonumber
\end{eqnarray}
where $h_0$ and $m_0$ are normalization constants.  The variable $R$
is nonnegative and measures the distance from the critical point in
the $(t,H)$ plane; it carries the power-law critical singularities.
The variable $\theta$ parametrizes the displacements along the line of
constant $R$.  The functions $m(\theta)$ and $h(\theta)$ are odd and
regular at $\theta=0$ and at $\theta=1$.  The constants $m_0$ and
$h_0$ can be chosen so that $m(\theta)=\theta+O(\theta^3)$ and
$h(\theta)=\theta+O(\theta^3)$.  The smallest positive zero of
$h(\theta)$, which should satisfy $\theta_0>1$, represents the
coexistence curve, i.e.\ $T<T_c$ and $H\to 0$.

The parametric representation satisfies the requirements of regularity
of the equation of state. Singularities can appear only at the
coexistence curve (due for example to the logarithms discussed in
Sec.~\ref{sec3b}), i.e.\ for $\theta=\theta_0$.  Notice that the
mapping (\ref{parrep}) is not invertible when its Jacobian vanishes,
which occurs when
\begin{equation}
Y(\theta) \equiv (1-\theta^2)m'(\theta) + 2\beta\theta m(\theta)=0.
\label{Yfunc}
\end{equation}
Thus the parametric representations based on the mapping (\ref{parrep})
are acceptable only if
$\theta_0<\theta_l$ where $\theta_l$ is the smallest positive 
zero of the function $Y(\theta)$.
One may easily verify that the asymptotic behavior (\ref{fxcc}) is 
reproduced simply by requiring that 
\begin{equation}
h(\theta)\approx \left( \theta_0 - \theta\right)^2 \qquad\qquad{\rm for}\qquad
\theta \rightarrow \theta_0.
\label{hcoex}
\end{equation}

The relation among the functions $m(\theta)$, $h(\theta)$ 
and $F(z)$ is given by
\begin{eqnarray}
&&z = \rho \,m(\theta) \left( 1 - \theta^2\right)^{-\beta},
\label{thzrel} \\
&&F(z(\theta)) = \rho \left( 1 - \theta^2 \right)^{-\beta\delta} h(\theta),
\label{hFrel}
\end{eqnarray}
where $\rho$ is a free parameter~\cite{GZ-97,CPRV-99}. Indeed, 
in the exact parametric equation the value of $\rho$ may be chosen
arbitrarily but, as we shall see, when adopting an approximation
procedure the dependence on $\rho$ is not eliminated.
In our approximation scheme we will fix $\rho$ to ensure the presence
of the Goldstone singularities  at the coexistence curve,
i.e.\ the asymptotic behavior (\ref{hcoex}).
Since $z=\rho\,\theta+O(\theta^3)$,
expanding  $m(\theta)$ and $h(\theta)$ in (odd) powers of $\theta$,
\begin{eqnarray}
m(\theta) &=& \theta  + \sum_{n=1} m_{2n+1}\theta^{2n+1} , \label{mhexp}\\
h(\theta) &=& \theta  + \sum_{n=1} h_{2n+1}\theta^{2n+1} ,\nonumber 
\end{eqnarray}
and using Eqs.~(\ref{thzrel}) and (\ref{hFrel}), 
one can find the relations among $\rho$, $m_{2n+1}$,
$h_{2n+1}$ and the coefficients $F_{2n+1}$ of the expansion of $F(z)$.

One may also write the scaling function $f(x)$ in terms of 
the parametric functions $m(\theta)$ and $h(\theta)$:
\begin{eqnarray}
&& x = {1 - \theta^2\over \theta_0^2 - 1} 
        \left[ {m(\theta_0)\over m(\theta)}\right]^{1/\beta} ,\label{fxmt}\\
&& f(x) = \left[ {m(\theta)\over m(1)}\right]^{-\delta} 
        {h(\theta)\over h(1)}.\nonumber
\end{eqnarray}

In App.\ \ref{univra} we report the definitions of some universal
ratios of amplitudes that have been introduced in the literature, and
the corresponding expressions in terms of the functions $m(\theta)$
and $h(\theta)$.

\subsection{Approximate polynomial representations}
\label{sec3d}

In order to construct approximate parametric representations we
consider polynomial approximations of $m(\theta)$ and $h(\theta)$.
This kind of approximation turned out to be effective in the case of
Ising-like sistems~\cite{GZ-97,CPRV-99}.  The major difference with
respect to the Ising case is the presence of the Goldstone
singularities at the coexistence curve. In order to take them into
account, at least in a simplified form which neglects the logarithms
found in Eq.~(\ref{fx3d}), we require the function $h(\theta)$ to have
a double zero at $\theta_0$ as in Eq.~(\ref{hcoex}).  Polynomial
schemes may in principle reconstruct also the logarithms, but of
course only in the limit of an infinite number of terms.

In order to check the accuracy of the results, it is useful to
introduce two distinct schemes of approximation.  In the first one,
which we denote as (A), $h(\theta)$ is a polynomial of fifth order
with a double zero at $\theta_0$, and $m(\theta)$ a polynomial of
order $(1+2n)$:
\begin{eqnarray}
{\rm scheme}\;\;\;({\rm A}):\qquad\qquad 
&&m(\theta) = \theta \left( 1 + \sum_{i=1}^n c_{i}\theta^{2i}\right) ,
        \label{scheme1}\\
&&h(\theta) = \theta \left( 1 - \theta^2/\theta_0^2 \right)^2. \nonumber
\end{eqnarray}
In the second scheme, denoted by (B), we set 
\begin{eqnarray}
{\rm scheme}\;\;\;({\rm B}):\qquad\qquad &&m(\theta) = \theta, 
        \label{scheme2}\\
&&h(\theta) = \theta \left( 1 - \theta^2/\theta_0^2 \right)^2
        \left( 1 + \sum_{i=1}^n c_{i}\theta^{2i}\right).
\end{eqnarray}
Here $h(\theta)$ is a polynomial of order $5+2n$ with a double zero at
$\theta_0$.  In both schemes the parameter $\rho$ is fixed by the
requirement (\ref{hcoex}), while $\theta_0$ and the $n$ coefficients
$c_{i}$ are determined by matching the small-field expansion of
$F(z)$.  This means that, for both schemes, in order to fix the $n$
coefficients $c_i$ we need to know $n+1$ values of $r_{2j}$, i.e.\ 
$r_6,...r_{6+2n}$.  Note that for the scheme (B)
\begin{equation}
Y(\theta) = 1 - \theta^2 + 2\beta\theta^2,
\label{Ytheta_def}
\end{equation}
independently of $n$,
so that $\theta_l = (1-2\beta)^{-1}$.
Concerning the scheme (A), we note that 
the analyticity of the thermodynamic quantities for $|\theta|<\theta_0$ 
requires the polynomial function $Y(\theta)$
not to have complex zeroes closer to origin than $\theta_0$.

In App.~\ref{app2} we present a more general discussion on the
parametric representations.

\subsection{Results}
\label{sec3e}

As input parameters for the determination of the parametric
representations, we use the best available estimates of the critical
exponents, which are $\alpha= - 0.01285(38)$ (from the experiment of
Ref.~\cite{LSNCI-96}), $\eta=0.0381(3)$ (from the high-temperature
analysis of Ref.~\cite{CPRV-00}).  Moreover we use the following
estimates of $r_{2j}$: $r_6=1.96(2)$ which is compatible with all the
estimates of $r_6$ reported in Table~\ref{summarygj}, and $r_8 =
1.40(15)$ which takes somehow into account the differences among the
various estimates.

\begin{table}
\caption{
Results for the parameters and the universal amplitude ratios using
the scheme (A), cf.\ Eq.~(\protect\ref{scheme1}), and the scheme (B),
cf.\ Eq.~(\protect\ref{scheme2}). The label above each column
indicates the scheme, the number of terms in the corresponding
polynomial, and the input parameters employed, beside the critical
exponents $\alpha$ and $\eta$.  Note that the quantities reported in
the first four lines do not have a physical meaning, but are related
to the particular parametric representation employed. Numbers marked
with an asterisk are inputs, not predictions.
}
\label{eqstresAB}
\begin{tabular}{lcccc}
\multicolumn{1}{c}{}&
\multicolumn{1}{c}{[(A) $n=1$; $r_6,r_8$]}&
\multicolumn{1}{c}{[(B) $n=1$; $r_6,r_8$]}&
\multicolumn{1}{c}{[(B) $n=2$; $r_6,r_8,r_{10}$]}&
\multicolumn{1}{c}{[(B) $n=2$; $r_6,r_8,A^+/A^-$]} \\
\tableline \hline
$\rho$       & 2.22(3)     & 2.07(2)   & 2.04(5)   & 2.01(4) \\
$\theta_0^2$ & 3.84(10)    & 2.97(10)  & 2.8(4)    & 2.5(2) \\
$c_1$        &$-$0.024(7)  & 0.28(3)   & 0.10(6)   & 0.15(5)  \\
$c_2$        &  0          & 0         & 0.01(2)   &  0.02(1) \\\hline
$r_{10}$     & $-$9.6(1.1) & $-$11(2)  &$^*-$13(7) & $-$7(5) \\
$A^+/A^-$    & 1.055(3)    & 1.057(2)  & 1.055(3)  & $^*$1.054(1) \\
$R_c$        & 0.123(8)    & 0.113(3)  & 0.118(7)  & 0.123(6) \\
$R_4$        & 7.5(3)      &  7.9(2)   & 7.8(3)    & 7.6(2) \\
$R_\xi^+$    & 0.353(3)    & 0.350(2)  & 0.352(3)  & 0.354(3) \\
$R_\chi$     & 1.38(9)     & 1.51(3)   & 1.47(8)   & 1.41(6) \\
$F_0^{\infty}$& 0.0303(3)  & 0.0301(3) & 0.0302(3) & 0.0304(3) \\
$c_f$        & 5(4)        & 62(41)    &           & 15(10) \\
\end{tabular}
\end{table}

The case $n=0$ of the two schemes (A) and (B) is the same, and
requires the knowledge of $\alpha$, $\eta$ and $r_6$. Unfortunately
this parametrization does not satisfy the consistency condition
$\theta_0^2<\theta_l^2 = (1-2\beta)^{-1}$. Both schemes give
acceptable approximations for $n=1$, using $r_8$ as an additional
input parameter. The numerical values of the relevant parameters and
the resulting estimates of universal amplitude ratios (see the
appendix for their definition) are shown in Table~\ref{eqstresAB}. The
errors reported are related to the errors of the input parameters
only. They do not take into account possible systematic errors due to
the approximate procedure we are employing.  We will return on this
point later.

\begin{figure}[tb]
\hspace{-1cm}
\vspace{0cm}
\centerline{\psfig{width=12truecm,angle=-90,file=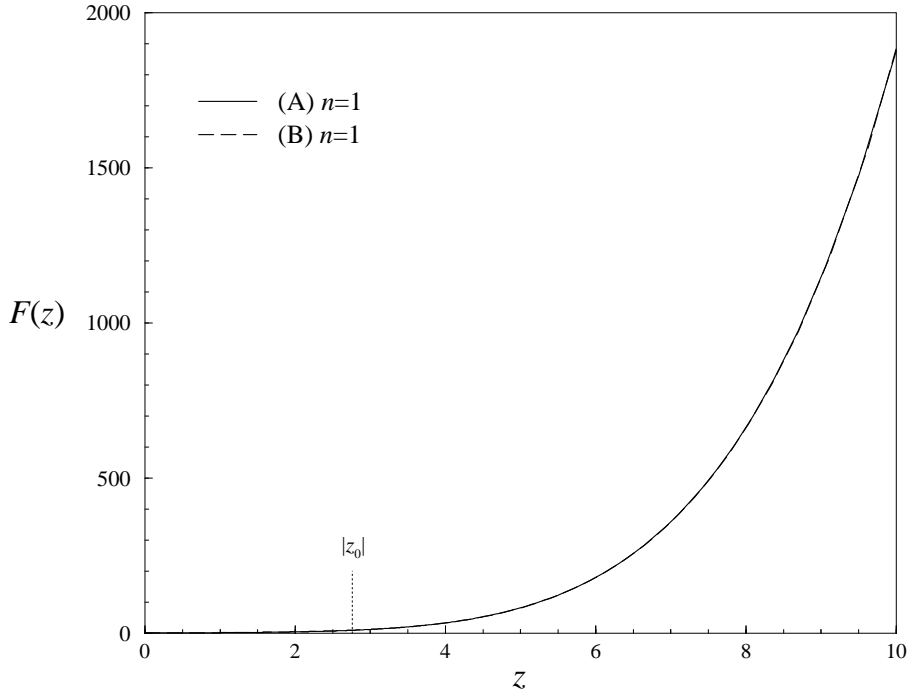}}
\vspace{0cm}
\caption{
The scaling function $F(z)$ vs.\ $z$. The convergence radius of the
small-$z$ expansion is expected to be $|z_0|=R_4^{1/2}\simeq 2.8$.}
\label{figFz}
\end{figure}

\begin{figure}[tb]
\hspace{-1cm}
\vspace{0.2cm}
\centerline{\psfig{width=12truecm,angle=-90,file=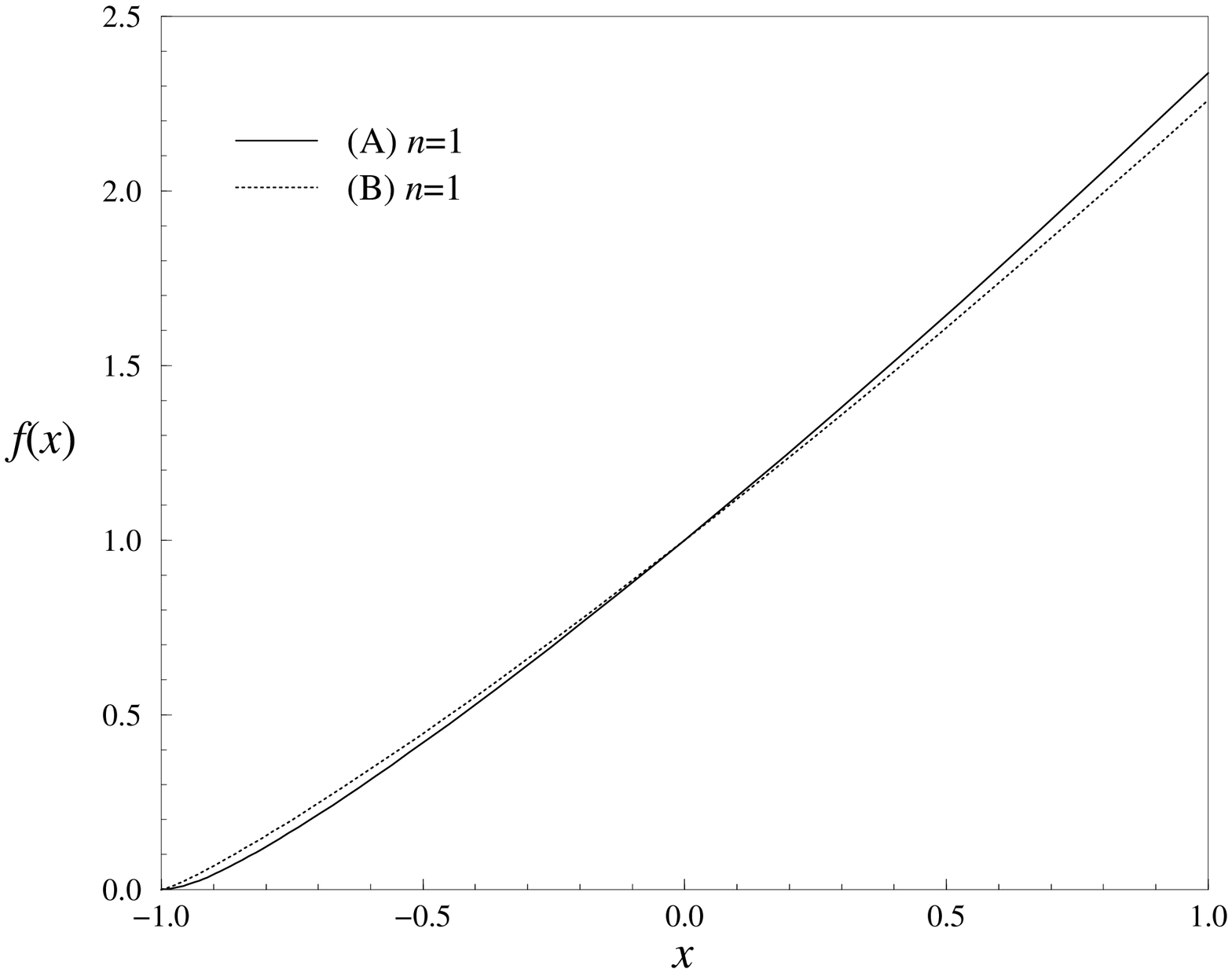}}
\vspace{-0.2cm}
\caption{
The scaling function $f(x)$ vs. $x$.
}
\label{figfx}
\end{figure}

In Figs.~\ref{figFz} and \ref{figfx} we show respectively the scaling
functions $F(z)$ and $f(x)$, as obtained from the approximate
representations given by the schemes (A) and (B) for $n=1$, using the
input values $\alpha=-0.01285$, $\eta=0.0381$, $r_6=1.96$ and
$r_8=1.4$.  The two approximations of $F(z)$ are practically
indinstinguishable in Fig.~\ref{figFz}.  This is also numerically
confirmed by the estimates of the universal costant $F^\infty_0$
(reported in Table~\ref{eqstresAB}), which is related to the large-$z$
behavior of $F(z)$:
\begin{equation}
F(z)\approx F_0^\infty z^\delta
\qquad\qquad{\rm for} \qquad z\rightarrow \infty.
\end{equation}
This agreement is not trivial since the small-$z$ expansion has a
finite convergence radius given by $|z_0|=R_4^{1/2}\simeq 2.8$.
Therefore, the determination of $F(z)$ on the whole positive real axis
from its small-$z$ expansion requires an analytic continuation, which
turns out to be effectively performed by the approximate parametric
representations we have considered.  We recall that the large-$z$
limit corresponds to the critical theory $t=0$, so that positive real
values of $z$ describe the high-temperature phase up to $t=0$.
Instead, larger differences between the approximations given by the
schemes (A) and (B) for $n=1$ appear in the scaling function $f(x)$,
especially in the region $x<0$ corresponding to $t<0$ (i.e.\ the
region which is not described by real values of $z$). Note that the
apparent differences for $x>0$ are essentially caused by the
normalization of $f(x)$, which is performed at the coexistence curve
$x=-1$ and at the critical point $x=0$ requiring $f(-1)=0$ and
$f(0)=1$.  Although the large-$x$ region corresponds to small $z$, the
difference between the two approximate schemes does not decrease in
the large-$x$ limit due to their slightly different estimates of
$R_\chi$ (see Table~\ref{eqstresAB}). Indeed, for large values of $x$,
$f(x)$ has an expansion of the form
\begin{equation}
f(x) = x^\gamma \sum_{n=0} f_n^\infty x^{-2n\beta}
\end{equation}
with $f_0^\infty=R_\chi^{-1}$.

We also considered the case $n=2$, using the estimate $r_{10}=-13(7)$.
In this case the scheme (A) was not particularly useful because it
turned out to be very sensitive to $r_{10}$, whose estimate has a
relatively large error.  Combining the consistency condition
$\theta_0<\theta_l$ (which excludes values of $r_{10}\lesssim -10$
when using the central values of $\alpha$, $\eta$, $r_6$ and $r_8$)
with the IHT estimate of $r_{10}$, we found a rather good result for
$A^+/A^-$, i.e.\ $A^+/A^-=1.053(4)$.  On the other hand, the results
for the other universal amplitude ratios considered, such as $R_\chi$,
$R_c$, $R_4$, etc..., although consistent, turned out to be much more
imprecise than those obtained for $n=1$, for example $R_\chi=1.2(3)$.
This fact may be explained noting that, except for the small interval
$-10\lesssim r_{10}\lesssim -9$ (this interval corresponds to the
central values of the other input parameters), the function
$Y(\theta)$, cf.\ Eq.~(\ref{Yfunc}), has zeroes in the the complex
plane which are closer to the origin than $\theta_0$.  Therefore, the
parametric function $g_2(\theta)$ related to the magnetic
susceptibility (see App.~\ref{univparrep}), and higher-order
derivatives of the free-energy, have poles within the disk
$|\theta|<\theta_0$.  On the other hand, in the case $n=1$,
$\theta_0^2$ was closer to the origin than the zeroes of $Y(\theta)$
for the whole range of values of the input parameters.

For these reasons, for $n=2$, we present results only for the scheme
(B).  Combining the consistency condition $\theta_0 < \theta_l$, which
restricts the acceptable values of $r_{10}$ (for example using the
central estimates of the other input parameters it excludes values
$|r_{10}|\gtrsim 12$), with the IHT estimate $r_{10}=-13(7)$, we
arrive at the results reported in Table~\ref{eqstresAB} (third column
of data).  The reported value has been obtained using 
$r_{10} \approx -9$.

The coefficients $c_i$, reported in Table~\ref{eqstresAB}, turn out to
be relatively small in both schemes, and decrease rapidly, supporting
our choice of the approximation schemes.  The results of the various
approximate parametric representations are in reasonable agreement.
Their comparison is useful to get an idea of the systematic error due
to the approximation schemes.  There is a very good agreement for
$A^+/A^-$, which is the experimentally most important quantity.  Our
final estimate is
\begin{equation}
A^+/A^- = 1.055(3).
\end{equation}
We mention that approximately one half of the error is due to the
uncertainty on the critical exponent $\alpha$, which unfortunately can
be hardly improved by present theoretical means.  The approximation
schemes (A) and (B) with $n=1$ provide independent results for
$r_{10}$, leading to the estimate $r_{10}=-10(3)$, which is agreement
with the IHT result $r_{10} = -13(7)$.  The determination of $c_f$,
cf.\ Eq.~(\ref{fxcc}), turns out to be rather unstable, indicating
that the approximate parametric representation we have constructed are
still relatively inaccurate in the region very close to the
coexistence curve.  The constant $c_f$ is very sensitive to the values
of the coefficients $r_{2j}$.  Improved estimates of $r_{2j}$ would be
important especially for $c_f$.

Finally, to further check our results, we applied again the scheme (B)
for $n=2$, replacing $r_{10}$ with the precise experimental estimate
$A^+/A^-=1.054(1)$ as input parameter.  In practice we fix the
coefficient $c_2$ in such a way to obtain the experimental estimate of
$A^+/A^-$.  The idea is to use the quantities known with the highest
precision to determine, within our scheme of approximation, the
equation of state and the corresponding universal amplitude ratios.
The results are reported in the last column of Table~\ref{eqstresAB}.

From the results of Table~\ref{eqstresAB} we arrive at the final estimates
\begin{eqnarray}
&&R_\xi^+\equiv (A^+)^{1/3} f^+ = 0.353(3) ,\\
&& R_c\equiv {\alpha A^+ C^+\over B^2} = 0.12(1) ,\\
&&R_\chi \equiv {C^+ B^{\delta-1}\over (\delta C^c)^\delta} = 1.4(1) ,\\
&&R_4\equiv  - {C_4^+ B^2\over (C^+)^3} = |z_0|^2 = 7.6(4) ,\\
&& F_0^\infty \equiv \lim_{z\rightarrow \infty} z^{-\delta} F(z) = 0.0303(3),\\
&&0 < c_f \lesssim 20.
\end{eqnarray}

\begin{table}[tbp]
\caption{
Estimates of universal amplitude ratios obtained in different approaches.
The $\epsilon$-expansion estimates of $R_c$ and $R_\chi$ have been
obtained by setting $\epsilon=1$ in the $O(\epsilon^2)$ series
calculated in Refs.\protect\cite{Rchi-e-exp}.
}
\label{summaryeqst}
\begin{tabular}{cccccc}
\multicolumn{1}{c}{}&
\multicolumn{1}{c}{IHT--PR}&
\multicolumn{1}{c}{HT, LT}&
\multicolumn{1}{c}{$d$=3 exp.}&
\multicolumn{1}{c}{$\epsilon$-exp.}&
\multicolumn{1}{c}{experiments}\\
\tableline \hline
$A^+/A^-$
     &1.055(3)&1.08 \cite{HAHS-76}& 1.056(4)\cite{LMSD-98}&1.029(13)\cite{Bervillier-86} & 1.054(1) \cite{LSNCI-96} \\
     &        &                   &                              &                              & 1.058(4) \cite{LC-83} \\
     &        &                   &                              &                              & 1.067(3) \cite{SA-84} \\
     &        &                   &                              &                              & 1.088(7) \cite{TW-82} \\\hline
$R_\xi^+$ 
& 0.353(3) & 0.361(4), 0.362(4) \cite{BC-99}&  0.3606(20)\cite{BB-85,BG-80}& 0.36 \cite{Bervillier-76}  & \\\hline

$R_c$    
&  0.12(1) &                      &  0.123(3) \cite{SLD-99}      &  0.106                           & \\ \hline

$R_\chi$  &  1.4(1) &                      &  &  1.407   & \\
\end{tabular}
\end{table}

In Table~\ref{summaryeqst} we compare our results with the available
estimates obtained from other theoretical approches and from
experiments (for a review see e.g. Ref.~\cite{PHA-91}).  Our precision
for $A^+/A^-$ is comparable with the estimate reported in
Ref.~\cite{LMSD-98}, obtained in the field-theoretic framework of the
minimal renormalization without $\epsilon$-expansion, which is a
perturbative expansion at fixed dimension $d=3$.  The agreement with
the experimental result of Ref.~\cite{LSNCI-96} is very good.

\subsection{Conclusions}
\label{sec3f}

Starting from the small-field expansion of the effective potential in
the high-temperature phase, we have constructed approximate
representations of the critical equation of state valid in the whole
critical region. We have considered two approximation schemes based on
polynomial representations that satisfy the general analytic
properties of the equation of state (Griffith's analyticity) and take
into account the Goldstone singularities at the coexistence curve.
The coefficients of the truncated polynomials are determined by
matching the small-field expansion in the high-temperature phase,
which has been studied by lattice high-temperature techniques.  The
schemes considered can be systematically improved by increasing the
order of the polynomials. However, such possibility is limited by the
number of known coefficients $r_{2j}$ of the small-field expansion of
the effective potential.  We have shown that the knowledge of the
first few $r_{2j}$ already leads to satisfactory results, for instance
for the specific-heat amplitude ratio.  Through the approximation
schemes we have presented in this paper, the determination of the
equation of state may be improved by a better determination of the
coefficients $r_{2j}$, which may be achieved by extending the
high-temperature expansion.  We hope to return on this issue in the
future.

Finally we mention that the approximation schemes which we have
proposed can be applied to other $N$-vector models. Physically
relevant values are $N=3$ and $N=4$. The case $N=3$ describes the
critical phenomena in isotropic ferromagnets~\cite{cubanis}.  The case
$N=4$ is interesting for high-energy physics: it should describe the
critical behavior of finite-temperature QCD with two flavours of
quarks at the chiral-symmetry restoring phase transition~\cite{PW-84}.

\appendix

\section{Universal ratios of amplitudes}
\label{univra}

\subsection{Notations}
\label{notationseqst}

Universal ratios of amplitudes characterize the critical behavior 
of thermodynamic quantities that do not depend on
the normalizations of the external (e.g.\ magnetic) field, order
parameter (e.g.\ magnetization) and temperature.  Amplitude ratios of
zero-momentum quantities can be derived from the critical equation of
state. We consider several amplitudes derived from the singular
behavior of the specific heat
\begin{equation}
C_H = A^{\pm} |t|^{-\alpha},
\label{sphamp}
\end{equation}
the magnetic susceptibility in the high-temperature phase
\begin{equation}
\chi = {1\over 2} C^{+} t^{-\gamma},
\label{chiamp}
\end{equation}
the zero-momentum four-point connected correlation function in the
high temperature phase
\begin{equation}
\chi_4 = {8\over 3} C_4^+ t^{-\gamma-2\beta\delta},
\label{chinamp}
\end{equation}
the second-moment correlation length in the high-temperature phase
\begin{equation}
\xi = f^{+} t^{-\nu},
\label{xiamp}
\end{equation}
and the spontaneous magnetization on the coexistence curve
\begin{equation}
M = B |t|^{\beta}.
\label{magamp}
\end{equation}
Using the above normalizations for the amplitudes, the zero-momentum
four-point coupling $g_4$, cf.\ Eq.~(\ref{grdef}), can be written as
\begin{equation}
g_4  = -{ C_4^+\over (C^+)^2 (f^+)^3}
\end{equation}
In addition,
one can also define amplitudes along the critical isotherm, such as 
\begin{equation}
\chi_L = C^c |H|^{-{\gamma/\beta\delta}}. \label{chicris}
\end{equation}

\subsection{Universal ratios of amplitudes from the parametric representation}
\label{univparrep}

In the following we report the expressions of the universal ratios of
amplitudes in terms of the parametric representation (\ref{parrep}) of
the critical equation of state.

The singular part of the free energy per unit volume can be written as
\begin{equation}
{\cal F} = h_0 m_0 R^{2-\alpha} g(\theta),
\end{equation}
where $g(\theta)$ is the solution of the first-order differential
equation
\begin{equation}
(1-\theta^2) g'(\theta) + 2(2-\alpha)\theta g(\theta) = Y(\theta) h(\theta)
\label{pp1}
\end{equation}
that is regular at $\theta=1$.
The function $Y(\theta)$ has been defined in Eq.~(\ref{Yfunc}).
The longitudinal magnetic susceptibility can be written as
\begin{equation}
\chi_L^{-1} = {h_0\over m_0} R^\gamma g_2(\theta),\qquad\qquad
g_2(\theta) = {2\beta\delta \theta h(\theta) + 
               (1-\theta^2) h'(\theta)\over Y(\theta)}.
\label{pp2}
\end{equation}
The function $g_2(\theta)$ must vanish at $\theta_0$ in order to reproduce the
predicted Goldstone singularities, according to
\begin{equation}
g_2(\theta)\sim \theta_0-\theta\qquad\qquad{\rm for}\qquad \theta \to \theta_0.
\end{equation}
From Eq.~(\ref{pp2}) we see that $g_2(\theta)$ satisfies this
condition if $h(\theta)\sim (\theta_0-\theta)^2$ for 
$\theta\to\theta_0$.

From the equation of state one can derive universal amplitude ratios
of quantities defined at zero momentum, i.e.\ integrated in the
volume. We consider
\begin{eqnarray}
&& A^+/A^- = 
(\theta_0^2 - 1 )^{2-\alpha} {g(0)\over g(\theta_0)},\\
&& R_c \equiv {\alpha A^+ C^+\over B^2} =
- \alpha (1-\alpha)(2-\alpha) (\theta_0^2 - 1 )^{2\beta} 
        [m(\theta_0)]^{-2} g(0),\\
&& R_4 \equiv  - {C_4^+ B^2\over (C^+)^3} = |z_0|^2 =
\rho^2 \,[m(\theta_0)]^2 \left(\theta_0^2-1\right)^{-2\beta},\\
&& R_\chi \equiv {C^+ B^{\delta-1}\over (\delta C^c)^\delta}=
(\theta_0^2-1)^{-\gamma} [m(\theta_0)]^{\delta-1} [m(1)]^{-\delta} h(1),
\end{eqnarray}

Using Eqs.\ (\ref{thzrel}) and (\ref{hFrel}) one can compute $F(z)$
and obtain the small-$z$ expansion coefficients of the effective
potential $r_{2j}$.  The constant $F_0^\infty$, which is related to
the behavior of $F(z)$ for $z\to \infty$, cf.\ Eq.~(\ref{fasymp}), is
given by
\begin{equation}
F^\infty_0 \equiv  \lim_{z\rightarrow \infty} z^{-\delta} F(z) = 
        \rho^{1-\delta} [m(1)]^{-\delta} h(1).
\end{equation}
Using the relations (\ref{fxmt}) concerning the scaling function
$f(x)$, one can easily obtain the constant $c_f$, which is related to
the behavior of $f(x)$ for $x\to -1$, cf.\ Eq.~(\ref{fxcc}),
\begin{equation}
 c_f \equiv \lim_{x\rightarrow -1} (1+x)^{-2} \,f(x). 
\end{equation}
We consider also the universal amplitude ratio 
$R_\xi^+\equiv (A^+)^{1/3} f^+$ which can be obtained from the
estimates of $R_4$, $R_c$ and $g_4$:
\begin{equation}
R_\xi^+ \equiv (A^+)^{1/3} f^+= \left( {R_4 R_c\over g_4}\right)^{1/3}.
\end{equation}
We mention that in the case of the superfluid helium
it is customary to define also the hyperuniversal combination 
\begin{equation}
R_\xi^{\rm T} \equiv (A^-)^{1/3} f^-_{\rm T} ,
\end{equation}
where $f^-_{\rm T}$ is the amplitude of a transverse correlation
length $\xi_{\rm T}$ defined from the stiffness constant $\rho_s$,
i.e.\ $\xi_{\rm T}=\rho_s^{-1}$.  $R_\xi^{\rm T}$ can be determined
directly from experiments below $T_c$ (see e.g. Ref.~\cite{PHA-91}).

\section{General discussion on the parametric representations}
\label{app2}

A wide family of parametric representations 
was introduced a long time ago
\cite{Schofield-69,SLH-69,Josephson-69}, in forms that can be related
to our Eqs. (\ref{parrep}).  Since the application of parametric
representations in practice requires some approximation scheme, one
may explore the freedom left in these representations and understand
how this freedom may be exploited in order to optimize the
approximation.  The parametric form of the equation of state forces
relations between the two functions $\overline{m}\equiv \rho
m(\theta)$ and $h(\theta)$, but it is easy to get convinced that one
of the two functions can be chosen arbitrarily. For definiteness,
let's take $\overline{m}$ to be arbitrary and find the constraints
that must be satisfied by $h(\theta)$ as a consequence of the equation
of state.

It is convenient for our purposes to establish these constraints by
imposing the formal independence of the function $F(z)$ from the
parametrization adopted for $\overline{m}\equiv \rho m(\theta)$, which
we may symbolically write in the form of a functional equation,
\begin{equation}
{\delta \over \delta \overline{m}} 
\left[ {\rho  h(\theta) \over (1-\theta^2)^{\beta \delta}}\right] = 0,
\end{equation}
keeping $z$ fixed.  By expanding $m(\theta)$ according to
Eq.~(\ref{mhexp}) and treating the coefficients $\rho$ and 
$c_{n} \equiv m_{2n+1}$ as variational parameters, we may turn the
above equation into a set of partial differential equations (keeping
$z$ fixed)
\begin{eqnarray}
&&{d \over d\rho}
\left[ {\rho  h(\theta) \over (1-\theta^2)^{\beta \delta}}\right] = 0 ,\\
&&{d \over d c_i}
\left[ {\rho  h(\theta) \over (1-\theta^2)^{\beta \delta}}\right] = 0 ,
\end{eqnarray}
which must be satisfied exactly for all $i$ by the function $h(\theta)$.
Simple manipulations lead to the following explicit form:
\begin{equation}
Y(\theta) \left( h + \rho {\partial h \over \partial \rho}\right) = 
m(\theta) \left[ (1- \theta^2) {\partial h \over \partial \theta} 
        + 2 \beta \delta \theta h\right],
\end{equation}
\begin{equation}
 Y(\theta) {\partial h \over \partial c_i} = 
\theta^{2i+1} \left[ (1- \theta^2) 
{\partial h \over \partial \theta} + 2 \beta \delta \theta h\right],
\end{equation}  
where $Y(\theta)$ is defined in Eq.\ (\ref{Ytheta_def}).
In turn, by expanding
\begin{equation}
h(\theta,\rho,c_i) = 
\theta + \sum_{n=1}^{\infty} h_{2n+1}(\rho, c_i) \theta^{2n+1},
\end{equation}
and substituting into the above equations one obtains an infinite set
of linear differential recursive equations for the coefficients
$h_{2n+1}$, which generalize the relations found in
Ref.~\cite{CPRV-99}, where the case $m(\theta)=\theta$ was analyzed.
A typical approximation to the exact parametric equation of state
amounts to a truncation of $h$ to a polynomial form.  We may in this
case refine the approximation by reinterpreting the first recursion
equations involving a coefficient $h_{2t+1}$ which is forcefully set
equal to zero as stationarity conditions, which force the parameters
$\rho$ and $c_i$ into the values minimizing the unwanted dependence of
the truncated $F(z)$ on the parameters themselves, i.e.\ on the choice
of the function $\overline{m}\equiv \rho m(\theta)$.  The above
procedure implies global (i.e.\ $\theta$-independent) stationarity,
and as a consequence all physical amplitudes turn out to be stationary
with respect to variations of $\rho$ and $c_i$.

These statements are fairly general, but it is certainly interesting
to consider the first few non trivial examples.  For the lowest order
truncation of $h$ we may adopt the parametrization
\begin{equation}
h(\theta) = \theta \left( 1- {\theta^2 \over \theta _0^2}\right)^p ,
\end{equation}
which includes both the Ising model in three dimensions ($p=1$) and
general $O(N)$ symmetric models with Goldstone bosons in $d$
dimensions ($p= 2/(d-2)$).  It is easy to recognize that the following
relationship must then hold:
\begin{equation}
{1 \over 6}\rho^2 + c_1 = \gamma - {p \over \theta_0^2}.
\end{equation}
Global stationarity implies that the stability conditions may be
extracted from the variation of any physical quantity. In particular
we may concentrate on the universal zero of $F(z)$, $z_0$, noting that
$z_0 = z(\theta_0)$, cf.\ Eq.~(\ref{thzrel}).  As a consequence of the
above results, the simplest models can all be described by the
parametrization
\begin{equation}
{z_0 \over \sqrt{6}} = {\sqrt{(\gamma-c_1)\theta_0^2 -p}
        \over (1- \theta_0^2)^{\beta}} (1+ c_1 \theta_0^2).
\end{equation}
Let us first consider the case $c_1=0$. The minimization procedure leads to
\begin{eqnarray}
&&\rho^2 = { 6 \gamma (\gamma- p) \over \gamma - 2 p \beta},\nonumber \\
&&\theta_0^2 = { \gamma - 2 p \beta \over (1 - 2 \beta)\gamma}.
\end{eqnarray}
Setting $p=1$ one immediately recognizes the linear parametric model
representation for the Ising model
\cite{Schofield-69,SLH-69,Josephson-69}.  Unfortunately when $p=2$ the
solution is not physically satisfactory, because it gives
$\theta_0^2<0$ for all $N\geq 2$, and therefore the scheme is useless
for models with Goldstone singularities.

Let us now include $c_1$. 
Requiring $z_0$ to be stationary with respect to variations of both parameters,
we obtain
\begin{eqnarray}
&&\rho^2 = { 6 \gamma (\gamma- p+1) \over 3 \gamma - 2 (p-1) \beta},
        \nonumber \\
&&\theta_0^2 = {3 \gamma - 2 (p-1) \beta \over (3- 2 \beta)\gamma},\nonumber \\
&&c_1 = \gamma {2(\gamma-p)+(2 \beta-1) \over 3 \gamma - 2 (p-1)\beta},
\end{eqnarray}
implying also
\begin{equation}
{|z_0| \over \sqrt{6}} = 
2 \left( {\gamma - p +1 \over 3 - 2 \beta}\right)^{{3 \over 2}- \beta} 
\left({\gamma \over 2 \beta}\right)^{\beta} .
\end{equation}
In the Ising model the above solution reduces to 
\begin{eqnarray}
&&\rho^2 = 2 \gamma, \\
&&\theta_0^2 = {3\over 3 - 2 \beta},\nonumber \\
&&c_1 = {2 \over 3}(\beta+\gamma)-1.\nonumber
\end{eqnarray}
Note that, substituting the physical values of the critical exponents
$\beta$ and $\gamma$ for the $N=1$ model, $c_1$ turns out to be a very
small number ($c_1 = 0.04256$) and the predicted numerical value of
$z_0$ is 2.8475, consistent within $1\%$ with the linear parametric
model prediction~\cite{CPRV-99}.  It is fair to say that in the Ising
case the above solution has a status which is comparable to the linear
parametric model, both conceptually and in terms of predicting power.
It is therefore possible to take it as the starting point of an
alternative approximation scheme whose higher-order truncations might
prove quite effective.

Unfortunately when we consider the $XY$ system, setting $p=2$ and
choosing the values of the exponents pertaining to $N=2$, the value of
$c_1$ becomes too large for the approximation to be sensible. Indeed
we get $c_1 = -0.6762$ and all testable predictions turn out to be far
away from the corresponding physical values.  It is however worth
exploring the features of this approach because, as we shall show, it
has formal properties which might prove useful when considering
parametric representations of the equation of state for higher values
of $N$.  Let us indeed consider the function $g_2(\theta)$ entering
the parametric representation of the magnetic susceptibility. We know
that this function will in general show singularities in the complex
$\theta$ plane corresponding to the zeroes of the function
$Y(\theta)$.  However, when substituting the expressions of
$h(\theta)$ and $m(\theta)$ obtained from the saddle-point evaluation
of the parameters $\rho$, $\theta_0$ and $c_1$, after some simple
manipulations, we find out that all singularities cancel and
\begin{equation}
g_2(\theta) = \left( 1- {\theta^2 \over \theta_0^2}\right)^{p-1} .
\end{equation}
This fact was already observed in the case $c_1=0$ for all values of
the truncation order $t$.  Therefore, the stationarity prescription is
a way to ensure a higher degree of regularity in the parametric
representation of thermodynamic functions.

Finally, let us observe that the stationary solution can be applied to 
the large-$N$ limit of $O(N)$ models in any 
dimension $ 2<d<4 $. In this limit
$\beta = {1 \over 2}$, $\gamma = p = {2/(d-2)}$.
As a consequence we obtain from our previous results
$\rho^2 = {12 /(d+2)}$, $\theta_0^2 = (d+2)/4$, $c_1 = 0$,
implying also
\begin{eqnarray}
&&h(\theta) = 
\theta \left(1- {4 \over d+2} \theta^2\right)^{2 \over d-2},\nonumber \\
&&g_2(\theta) =  \left(1- {4 \over d+2} \theta^2\right)^{4-d \over d-2}.
\end{eqnarray}
We therefore obtained an exact parametrization of the equation of
state in the large-$N$ limit for all $d$. Thus, for sufficiently large
values of $N$, the scheme we have defined may be a sensible starting
point for the parametric representation of the thermodynamical
functions in the critical domain.


\begin{thebibliography}{99}
\bibitem{footnote_singularity}
The observed agreement is also a good indication of the correctness 
of the analysis of the experimental data, questioned in Refs.
\cite{Arp,Fliessbach}, and
excludes different approaches that predict a logarithmic singularity
\cite{Fliessbach}.
\bibitem{cflargeN}
Using the formulae of Ref.~\cite{BW-73}, one 
can evaluate the constant $c_f$ in the large-$N$ limit finding
$c_f = 1 + 1.64657/N + O(1/N^2)$.
\bibitem{epsexp} For $N$-vector models with $N>1$, the
$\epsilon$-expansion of the whole equation of state is known to
$O(\epsilon^2)$~\cite{BWW-72}; instead, its small-field expansion in
the symmetric phase has been calculated to
$O(\epsilon^3)$~\cite{PV-00}.
\bibitem{cubanis}
Note that the isotropic Heisenberg fixed point turns out to be
unstable in the presence of cubic anisotropy~\cite{PV-00b}, which is a
general feature of real ferromagnets due to the lattice structure.
Indeed cubic anitropy drives the renormalization-flow toward a stable
fixed-point characterized by the discrete cubic symmetry. However, if
the cubic interactions are weak, the isotropic Heisenberg model may
still describe the critical behavior of the system in a relatively
wide region near the critical point.  Of course, whatever is the
strength of the cubic interaction, there will be a crossover to the
cubic fixed point for $T\to T_c$. But, if the interaction is weak,
this occurs only very close to $T_c$.
\end{thebibliography}
\end{document}